\documentclass[a4paper]{jpconf}

\newcommand{\DPV}{\Delta \partial V}
\newcommand{\Sp}{S_{\partial 1}}
\newcommand{\SpD}{S_{\partial 1 \ \DPV}}
\newcommand{\intc}[3]{\int_{#1} d^{#2}x \ #3}
\newcommand{\intct}[3]{\int_{#1} d^{#2}\tilde{x} \ #3}
\newcommand{\sqn}[1]{\sqrt{|#1|}}
\newcommand{\Lm}{\mathcal{L}}
\newcommand{\gah}[1]{\gamma^{#1}}
\newcommand{\gal}[1]{\gamma_{#1}}
\newcommand{\cfac}[1]{\frac{c^4}{#1 \pi G}}
\newcommand{\cface}[1]{\frac{c^4 \epsa{}}{#1 \pi G}}
\newcommand{\xpara}{x^{||}}

\newcommand{\nline}{\nonumber \\ }
\newcommand{\bu}{\bar{u}}
\newcommand{\bv}{\bar{v}}

\newcommand{\epsa}[1]{\varepsilon_A \ #1}
\newcommand{\inte}[2]{\epsa{\int_{#1} d^3x \ #2}}

\newcommand{\Jac}[2]{\big| \frac{e^2_{#1}}{e^2_{#2}}  \big|}
\newcommand{\sumk}{\sum_{k=1}^N}
\newcommand{\tT}[1]{\tilde{T}_{#1}}
\newcommand{\half}{\frac{1}{2}}

\newcommand{\frama}[1]{e^{#1 \ I}_a}
\newcommand{\framb}[1]{e^{#1 \ J}_b}
\newcommand{\dframa}[1]{d^I_a(#1)}
\newcommand{\dframb}[1]{d^J_b(#1)}
\newcommand{\gfram}{(\frama{vac}+\dframa{\tau}) \eta_{IJ} (\framb{vac}+\dframb{\tau})}
\newcommand{\gframv}{\frama{vac} \eta_{IJ} \framb{vac}}
\newcommand{\Gam}[1]{\Gamma_{#1}}

\newcommand{\SpExpr}{\cfac{8} \sum_A \inte{B_A}{\sqrt{|\gamma|} \ K}}
\newcommand{\NdSEq}{N^c_{ab} \ d^3\Sigma_c = \epsa{N^{ab} \ d^3x}}

\begin{document}
\title{Quantum matter characteristics from multiple Rindler observer formulation of general relativity}

\author{P A Mandrin}

\address{IMSD, Lavaterstrasse~103, 8002~Z\"urich, Switzerland}

\ead{pierre.mandrin@imsd.ch}

\begin{abstract}
By considering matter as a constraint on the availability of gravitational degrees of freedom and accounting for the statistical interpretation of Rindler horizons, the freedom to construct quantum gravity theories reproducing General Relativity and Quantum Field Theory (QFT) as special cases is considerably reduced. On one hand, the mathematical structure of quantum gravity is restricted by the properties of Quantum Mechanics. On the other hand, one can predict a value for the fundamental quantum constant of gravity which is related to the Planck area via the Planck constant. These findings are compatible with spin-less particles of matter. In the context of canonical ensemble statistics, the von Neumann entropy concept is found to extend from matter to gravity.
An important motivation and pillar for this development is the concept of
multiple observer statistics and the total entropy perceived by Rindler observers with a particular spacing, which has a one-to-one correspondence to the Gibbons-Hawking-York boundary term.

\end{abstract}

\section{Introduction}
\label{intro}

So far, experimental evidence is lacking of which of the approaches to quantum gravity (if any) should be applicable. Because of its smallness, the value of the fundamental constant of quantum gravity is also unknown. However, there might be some profound relation between gravity and matter. This possibility has received little attention in the past, but could help reduce the realm of admissible theories. We shall consider here matter as a constraint which restricts the gravitational degrees of freedom. The goal of this article is to focus on a gravity-matter relation instead of quantising some set of canonical variables. 

As a skeleton of our formalism, we use the ideas of black hole thermodynamics \cite{Bekenstein,Hawking,Wald} which have been extended to Rindler space-times, where the Unruh radiation \cite{Unruh} corresponds to the Hawking radiation of black holes. In the limit of stationarity, Einstein's field equations appear as a manifestation of the First Law applied to the null-horizon \cite{Jacobson,Padmanabhan_2014}, so that any null-surface acquires a local, observer-dependent temperature $T = \kappa / (2\pi)$ with surface gravity $\kappa$, and an entropy $S_{BHJP}=A/(4 L_p^2)$, where $A$ is the horizon surface area and $L_p$ is the Planck length.

This concept has recently been extended to describe multiple Rindler observers as required for observing a horizon temperature, this yields a Boltzmann-like 3d-space entropy -- we call it \mbox{m-entropy} $S$ -- which is equivalent (up to signatures) to the Gibbons-Hawking-York boundary term \cite{York,GH} if the density of observers is proportional to the external curvature \cite{Mandrin_2017}. In contrast to $S_{BHJP}$, the existence of $S$ does not require the underlying microscopic model to be in local equilibrium, since $S$ is not paired with a temperature. This allows for hypothetical departures from General Relativity and opens a door to quantum behaviour on a space of reduced dimension ($2+1$), in correspondence to the holographic principle \cite{Dim_Red}.  

The present article investigates departures from the vacuum geometry. These departures are interpreted in terms of matter fields, the latter play the role of constraints. We derive some formal properties of matter (in the context of GR, we only consider spin-0 fields) and investigate the resulting fundamental quanta of gravity.

For the convenience of the reader, we also summarise in this article: 

\begin{itemize}
\item the concept of m-entropy which has been formerly derived using  \cite{Padmanabhan_2014},
\item how the Gibbons-Hawking boundary term can be subject to the variation principle without involving the bulk action (cf \ref{boundary_action}) and
\item the multiple observer description of gravity (cf \ref{thermodynamics}); its statistical interpretation is summarized in \ref{sec:statistics}.
\end{itemize}

For the quantum properties of the m-entropy, we refer to \cite{Mandrin_2017}, where we have sketched the derivation of quantum uncertainties in a system with a small number of quantum bits of gravity, as well as the derivation of some typical transition probabilities between quantum states.

\section{Synopsis of the multiple observer formulation of GR}
\label{multilayer}

The starting point is the Gibbons-Hawking boundary term for a compact 4-volume $V$; we omit any terms free of derivatives of the metric for simplicity, and no matter term is added:

\begin{equation}
\label{S_d1_syn}
\Sp = \SpExpr.
\end{equation}

\noindent In (\ref{S_d1_syn}), the integration is performed over the piece-wise smooth boundary $\partial V = \sum_A B_A$ of $V$, where every component $B_A$ is time- or space-like. Furthermore, $K = K_{ab} \gamma^{ab}$, where $\gamma_{ab}$ is the induced metric on $\partial V$ with determinant $\gamma$, $K_{ab}$ is the exterior curvature on $\partial V$ with unit normal vector $n^a$, namely $K_{ab} = -\frac{1}{2} \Lm_\perp \gamma_{ab} = -g^{cd}\gamma_{da} \nabla_c n_b$, $\Lm_\perp=\Lm_n$, and $\varepsilon_A = n^an_a$ on $B_A$. Equivalently, following the notation of \cite{Padmanabhan_2014}, (\ref{S_d1_syn}) can be written as

\begin{equation}
\label{S_d1_P_syn}
\Sp = \cfac{16} \int_{\partial V} N^c_{ab} \ f^{ab} \ d^3\Sigma_c, \qquad f^{ab} = \sqrt{-g} \ g^{ab}, \qquad \NdSEq,
\end{equation}

\noindent where $N^c_{ab}$ is defined as the momentum conjugate to $f^{ab}$, $d^3\Sigma_c$ is the the 3-surface element covector and $N_{ab} = K_{ab} - K \gamma_{ab}$. $\Sp$ can be subject to the variation principle (without involving the bulk, cf \ref{boundary_action}). We divide the boundary $\partial V$ into rectangular cuboids $\DPV$ with edges short enough for geometric variations to be negligible across $\DPV$ and large enough for quantum gravity effects to be negligible. For the portion of boundary term (\ref{S_d1_syn}) contained within $\DPV$, it can be shown (cf \ref{thermodynamics}) \cite{Mandrin_2017,Majhi_Padmanabhan}:

\begin{eqnarray}
 \SpD  & = &  \cface{32} \intc{\DPV}{3}{\sqn{\gamma} \ \gah{ab} \ \Lm_\perp \gal{ab}} \nline
& = &  \cface{32} \sumk \intct{A(\xpara_k)}{2}{\bigg[ \sqn{\gal{(u)}}\Jac{||}{u} \ \Delta u \ \gah{ab}_{(u)} \ \Lm_v \gal{ab}^{(u)} + \sqn{\gal{(v)}}\Jac{||}{v} \ \Delta v \ \gah{ab}_{(v)} \ \Lm_u \gal{ab}^{(v)} \bigg]} \nline
\label{general_null_syn}
& = &  \cface{32} \sumk \sum_{w=u,v} \int_{\Delta w} \intct{A(\xpara_k)}{3}{\sqn{\gal{(w)}}\Jac{||}{w} \ \gah{ab}_{(w)} \ \Lm_v \gal{ab}^{(w)}}  \\
\label{multi_observers_syn}
& = &  \sumk \bigg[ \int_{\Sigma_{(u)}(\xpara_k)} \tT{u} \ s \ du \ d^2x + \int_{\Sigma_{(v)}(\xpara_k)} \tT{v} \ s \ dv \ d^2x \bigg]  ,
\end{eqnarray}

\noindent where $e^J_a$ is the induced triad on $\partial V$, $\gal{ab}^{(w)}$ and $\gah{ab}_{(w)}$ are the metric and inverse metric induced on the hypersurface $w =$ constant, respectively, $w = u,v$ are null coordinates, $\tilde{x}^\mu = (u,v,x^2,x^3)$, $\tT{w} = \Jac{||}{w} \ T_w$ and $T_w$ is the horizon temperature observed on the hypersurface $w =$ constant. If we choose precisely $\Delta w = \tT{w}^{-1}$, the temperatures disappear together with the null distances, i.e. $\SpD$ can be interpreted as a Boltzmann-like entropy if the spacings between the observers are wright at the limit of the classical spatial resolution, i.e. the density of observers is proportional to the resolvable normal change of $\gamma_{ab}$. Since $\SpD$ is not time-dependent, it does not depend on any thermalization condition, but exists beyond stationarity. Only in the case of stationarity does $\SpD$ reduce to a sum of products of horizon entropies and temperatures. From the point of view of the Rindler observers, the total entropy within $\DPV$ is the m-entropy, i.e. the sum of the entropies of all the null-strips contained in (\ref{multi_observers_syn}) -- this is the lack of information related to a thermodynamic measurement process which takes some time and/or distance.

As a consequence, the boundary space naturally splits into multiple layers of 2d-surfaces (like a foliation across a given direction), and every layer consists of a collection of ordered surface microstates $|q\rangle$ (cf \ref{sec:statistics}). The layer structure is similar to a computer hard disk, where every layer stores a certain number of bits. For gravity, however, there is no a-priori restriction about which fraction of the states is on which layer. A stationary point of $\SpD$ is a macroscopic geometry, for which the number of microstate configurations is (locally) maximum.

Multiple observer statistical gravity is not the first development for which some kind of analogy between the gravitational action and the notion of ''entropy'' is suggested \cite{Mandrin1, Mandrin2, Munkhammar}. However, it is the first derivation of such a relation on the basis of horizon thermodynamics. For quantum mechanics, such an analogy has been conjectured even earlier \cite{Lisi}.

\section{Quasi-stationarity condition as a matter generating mechanism}
\label{quasi_stationarity}

Although the stationary point of $\Sp$ is the geometry of vacuum ($\gamma_{ab}^{vac}$), there is enough room for non-vacuum metrics since the fraction of vacuum ($N_K, \{n_{Ki}|i\le N_K\}$)-states is small against the non-vacuum states. Our goal is:
\begin{enumerate}
\item constrain  $\gal{ab}$ to non-vacuum configurations with fixed ``distance'' from $\gamma_{ab}^{vac}$ and
\item solve for a stationary point $\delta \Sp\big|_{constr}=0$ under the constraint (i) (quasi-stationarity).
\end{enumerate}
Procedure: Be $\{\Gam{\dframa{\tau}}\}$ with $\Gam{\dframa{\tau}} = \gfram$ a family of distinct curves in the neighbourhood of $\gal{ab}^{vac} = \gframv$ in the space of $\gal{ab}$ 
with $0 \le \tau \le \epsilon$, $\Gam{\dframa{0}} = \gamma_{ab}^{vac}$ and $\Gam{\dframa{\epsilon}} \ne \gal{ab}^{vac}$. We impose the stationarity condition $\delta (S\big|_{\gamma_{ab}=\Gam{\dframa{\epsilon}}}) = 0$, where $\dframa{\epsilon}$ is varied, up to gauge transformations of $\gal{ab}$ leaving $S$ invariant. To impose condition (i), we introduce a compensation term $S_m(\Gam{\dframa{\epsilon}},\varphi)$ such that $\delta (S + S_m) = 0$, and the independent object $\varphi$ ensures a fixed ``distance'' between $\gal{ab}$ and $\gal{ab}^{vac}$. We call this the ``compensation method''. We look for a general ansatz for $S_m$. A special case with simplified treatment is

\begin{equation}
\label{eq:tenuous}
|\Gam{\dframa{\epsilon}} - \gamma_{ab}^{vac}| / |\gamma_{ab}^{vac}| \ll 1,
\end{equation}

\noindent which we call the tenuous limit, where $|\gamma_{ab}|$ denotes the largest absolute component of $\gamma_{ab}$.

\subsection{Tenuous limit}
\label{tenuous}

We expand the induced metric up to the first order in the real ``distance`` parameter $\tau$:

\begin{eqnarray}
\gamma_{ab}(\tau) & \approx & \eta_{ab} + \zeta_{ab} \tau, \nonumber \\
\label{eq:gamma_o_lambda}
\gamma^{ab}(\tau) & \approx & \eta^{ab} - \bar{\zeta}^{ab} \tau,
\end{eqnarray}

\noindent where we define $\bar{\zeta}^{ad} = \eta^{ab}\zeta_{bc}\eta^{cd}$ so that $\gah{ab}\gal{bc} = \delta^a_c$ holds to first order in $\tau$. Next, we write the integrand $\sigma_g$ of the boundary term (\ref{S_d1_syn}) to lowest order in $\tau$:

\begin{equation}
\label{eq:sigma_g}
\sigma_g = \cface{16}\sqrt{|\gamma|}\gah{ab}\Lm_\perp\gal{ab} \approx -\cface{16}\sqrt{|\gamma|}\bar{\zeta}^{ab}\Lm_\perp\zeta_{ab} \tau^2.
\end{equation}

\noindent Notice that the first order term in $\tau$ vanishes because $\gal{ab}(0)$ is a stationary point for $S$. Because $\sigma_g$ is of second order in $\tau$, so must be the compensation term integrand $\sigma_m$. We reduce (\ref{eq:sigma_g}) to  

\begin{equation}
\label{eq:zeta_red}
\sigma_g \approx -\cface{32}\sqrt{|\gamma|}\Lm_\perp(\bar{\zeta}^{ab}\zeta_{ab}) \tau^2.
\end{equation}

\noindent In the expression for (\ref{eq:zeta_red}), we replace $-\bar{\zeta}^{ab}\zeta_{ab} \tau^2$ by the square of a constraining scalar function $\varphi \sim  \tau$ to construct $\sigma_m$:

\begin{equation}
\label{eq:Lm}
\sigma_m \approx \vartheta\sqrt{-g}\nabla_\perp\varphi^2 = 2\vartheta\sqrt{-g}\varphi\nabla_\perp\varphi,
\end{equation}

\noindent where $\vartheta$ is an arbitrary constant (still to be determined). A form like (\ref{eq:Lm}) generates gravity-independent dynamical degrees of freedom: a field. Choosing at least $\sigma_m \sim \nabla_\perp\varphi^2$ rather than $\sim \nabla_\perp\varphi$ is necessary in order that  a non-trivial bulk Lagrangian for $\varphi$ with boundary term $S_m$ and non-trivial equations of motion exists (we ignore theories without a Lagrangian). More generally, $\varphi$ can be allowed to be complex, $\sigma_m \sim \nabla_\perp\varphi^*\varphi$. Notice that the objects $\bar{\zeta}^{ab}$ and $\zeta_{ab}$ at the stationary points of $S + S_m$ ($\tau=\epsilon$ fixed) do not depend on the choice of $\epsilon$.

Because our starting point was the boundary term for GR, we expect that $\varphi$ is a spin-less matter field, namely the Klein Gordon (KG) field or a some transformation of it. To examine this, we compute the boundary KG term $S_{bKG}$ from the bulk KG Lagrangian $\Lm_{KG}$ or action $S_{KG}$ (restricting ourselves to a real $\varphi$), and use Gauss' Theoreme:

\begin{eqnarray}
\Lm_{KG} & = & \frac{1}{2} (\partial^\mu \varphi \partial_\mu \varphi - \frac{m^2 c^2}{\hbar^2} \varphi^2), \nonumber \\
\label{eq:dS_KG}
\delta S_{KG}/\delta \varphi & = & \int_V d^4x \sqrt{-g} (- \nabla^\mu\nabla_\mu - \frac{m^2 c^2}{\hbar^2})\varphi + \delta S_{bKG} / \delta\varphi, \\
\label{eq:Sb_KG}
S_{bKG} & = & \int_V d^4x \sqrt{-g} \nabla^\mu(\varphi\nabla_\mu\varphi) = \int_{\partial V} d^3x \sqn{\gamma} \varphi\nabla_\perp\varphi.
\end{eqnarray}

\noindent We see that $S_m$ and $S_{bKG}$ are of the same mathematical form, so that they compensate each other (and the Klein Gordon Equation holds) if $\varphi$ in (\ref{eq:Lm}) equals $\varphi$ in (\ref{eq:Sb_KG}) and $\vartheta = -\frac{1}{2}$.

\subsection{Higher density}
\label{dense}

If (\ref{eq:tenuous}) is not satisfied (higher density regime), it follows from the non-linearity of $S_g$ that a metric description similar to (\ref{eq:gamma_o_lambda}) would cause $\zeta_{ab}$ at the stationary points (while $\tau$ is fixed) to depend on $\tau$ in a more complex way:

\begin{eqnarray}
\gamma_{ab}(\tau) & \approx & \eta_{ab} + \zeta_{ab}(\tau), \nonumber \\
\label{eq:gamma_f_lambda}
\gamma^{ab}(\tau) & \approx & \eta^{ab} - \bar{\zeta}^{ab}(\tau),
\end{eqnarray}

\noindent where $\bar{\zeta}^{ab}(\eta_{bc}+\zeta_{bc}) =  \eta^{ab} \zeta_{bc}$, namely, the tensorial structure of $\zeta_{ab}$ itself would depend on $\tau$ as well. Nevertheless, we can still proceed in analogy to the tenuous limit, as for (\ref{eq:zeta_red},\ref{eq:Lm}).

\section{Quanta of matter, quanta of gravity, Planck area and canonical ensemble}
\label{quanta}

Consider a cuboid $\DPV$ (on $\partial V$) which is foliated across $x^K$ ($K$ is a Minkowski index) into $N_K$ pieces of 2d-surfaces of aria $A_K$ representing $n_K$ bits each, where $N_K$ is defined according to \ref{sec:statistics}. If $n_q$ defines the number of possible states of every bit and if every microstate of $\DPV$ is given by a pair ($N_K, n_K$), the m-entropy of $\DPV$ is \cite{Mandrin_2017} (see also \ref{sec:statistics})

\begin{eqnarray}
\label{eq:s_K}
s_K & = & N_K n_K \ln{n_q}, \\
\label{eq:s_K_i}
s_K & = & \sum_{i} n_{Ki} \ln{n_q}.
\end{eqnarray}

\noindent ($K$ is not used as a tensorial index in (\ref{eq:s_K})). (\ref{eq:s_K}) generalizes to (\ref{eq:s_K_i}) if we allow the number $n_{Ki}$ of bits on every $i$th surface to vary individually.
Equations (\ref{eq:s_K},\ref{eq:s_K_i}) describe the ordered quantum bit structure for GR. We analyse now how (\ref{eq:s_K_i}) can be related to quantum matter in the tenuous limit. Consider a curve $\Gam{\dframa{\tau}}$  in $\gal{ab}$-space, made of points satisfying the stationarity condition $\delta (S\big|_{\gamma_{ab}=\Gam{\dframa{\tau}}}) = 0$ (for $\tau$ fixed). We change $s_K/(\ln{n_q})$ along $\Gam{\dframa{\tau}}$ by a number of bits $\Delta(\sum_{i} n_{Ki})$ so that the change of $\gal{ab}$ is no less than the smallest classically observable difference. To maintain stationarity, we modify $\varphi = \varphi(\tau)$ (by the same number of quanta as bits) via

\begin{equation}
\label{eq:D_change}
\Delta (N_K n_K) \ln{n_q} = \frac{d\sigma_g}{d\tau} \Delta\tau = - \frac{d\sigma_m}{d\tau} \Delta\tau \approx - \frac{\partial\sigma_m}{\partial\varphi} \Delta\varphi.
\end{equation}

\noindent In (\ref{eq:D_change}), the stationarity condition has been required to hold in physical equivalence to the quantum mechanical commutation relations which reflect the equations of motion. Subjecting $\varphi$ to second quantisation, $\varphi \rightarrow \hat{\varphi}$, converts (\ref{eq:D_change}) into an operator equation (i.e. $\gal{ab}$ is not sharp)

\begin{equation}
\label{eq:Q_change}
\frac{d\sigma_g}{d\tau} \hat{\tau} = - \frac{\partial\sigma_m}{\partial\varphi} \hat{\varphi}, \qquad \hat{\tau} = \sum_i \tau_{i-} a^\dagger_i + \tau_{i+} a_i,
\end{equation}

\noindent with $a_i|n_{Ki}\rangle=n_{Ki}^{1/2}|n_{Ki}-1\rangle$, $a^\dagger_i|n_{Ki}\rangle=(n_{Ki}+1)^{1/2}|n_{Ki}+1\rangle$. (\ref{eq:D_change}) establishes how every quantum change of gravity yields a quantum change in matter. Conversely, $\hbar$ fixes the fundamental quantum constant of gravity, using (\ref{eq:Lm}, \ref{eq:Sb_KG}, \ref{eq:D_change}) and $\int_{\partial V} d^3x \sqn{\gamma} \varphi\nabla_\perp\varphi = \hbar c$:

\begin{equation}
\label{eq:quantum_grav}
\frac{1}{32\pi}\bigg|\int_{\partial V} d^3x \sqrt{|\gamma|}\Lm_\perp(\bar{\zeta}^{ab}\zeta_{ab}) \Delta(\tau^2) \bigg| = \frac{\hbar G}{c^3} = L_p^2,
\end{equation}

\noindent which relates the quantum of ``integrated surface gravity'' to the Planck area. Notice that $L_p^2$ is not just an ad-hoc scale of breakdown of GR, but fixes the elementary quantum constant.

Equation (\ref{eq:quantum_grav}) simplifies further if we consider the example of a scalar particle in the form of a monochromatic planar wave function with wave covector $k_\mu = (k_0,k_\perp,0,0)$. The non-zero elements of $(\zeta_{ab}\tau)$ also yield a monochromatic plane wave function. 
If we choose $V$ to be a hypercuboid with edges parallel to the coordinate axes and the two space-like hypersurfaces $\Sigma_{\pm}$ separated by $\pi/k_0$ (half a period) and the time-like hypersurfaces $\Pi_{\pm}$ perpendicular to $(0,k_\perp,0,0)$  very near to each other (distance $\ll \pi/k_\perp$), we obtain

\begin{equation}
\label{eq:quantum_const}
\big|\bar{\zeta}^{ab}\zeta_{ab}\big|_{max} \Delta(\tau^2) = 16\pi \frac{L_p^2}{|\Sigma_+| k_0},
\end{equation}

\noindent where $|\Sigma_+|$ is the 3-volume of $\Sigma_+$, since only $\Sigma_{\pm}$ contribute significantly. 

Finally, a cuboid $\DPV$ can be considered to be immersed in a ``bath'' of matter field ($\varphi$). It can therefore be described quantum mechanically as a canonical ensemble. 
The ensemble average entropy $s_N$ of all possible multiple observer states of $\DPV$, with bit-number eigenstates $|n_{lKi}\rangle$, density matrix $\hat{\rho}$ and matter compensation term $s_m(\DPV,\varphi)$, is the von Neumann entropy:

\begin{eqnarray}
\label{s_N_mixed}
s_N & = & \tr{(\hat{\rho} \ln{\hat{\rho}})}, \\
\label{rho}
\rho_{lm} & = & |n_{lKi}\rangle \ \frac{\exp{[s_K(n_{lKi})+s_m]}}{\sum_k \exp{[s_K(n_{kKi})+s_m]}} \ \bigg[\prod_j\delta_{n_{lKj}n_{mKj}}\bigg] \ \langle n_{mKi}| \quad (i\ge 1).
\end{eqnarray}

\section{Conclusions}
\label{conclusions}

By constraining the Gibbons-Hawking-York boundary term to lowest physically nontrivial order (quadratic) in the constraint ``parameter'' $\varphi$ while taking the first derivative over to the constraint term, we have obtained the emergence of a spin-less matter field (which is compatible with the torsionlesness of GR), at least in the tenuous limit. This fact suggests a profound relation between gravity and matter. Constraining the boundary term (and not the bulk action) is largely motivated by the multiple observer statistical interpretation of GR. 
We infer that quantum matter might be intimately related to the quantum structure of gravity as well, so that the full quantum mechanical treatment can be transferred from quantum matter to the gravitational field. 
Under this condition, the fundamental area of one bit of ``integrated surface gravity'' has been found to be $32\pi$ times the Planck area. If we take this computation literally, that would mean that the quantisation of gravity has indirectly already been quantified since $\hbar$ has been determined. Finally, von Neumann's quantum canonical ensemble statistics straight-forwardly extends to the ``matter bath'' description of quantum gravity. Since our approach does not rely on the quantisation prescription of an a priori model of canonically conjugated variables, but instead on the concept of ``gravity constrained by matter'' with a generalised statistical mechanical background, the applicability of our investigations might be quite general.

\appendix
\section{The boundary term and the variation principle}
\label{boundary_action}

For a compact 4d-region $V$ with boundary $\partial V$, consider the Hilbert-Einstein action $S_H$ and the Gibbons-Hawking-York boundary term $\Sp$, with the symbols defined as in Section \ref{multilayer},

\begin{eqnarray}
\label{S_H}
S_H & = & \cfac{16} \intc{V}{4}{\sqrt{-g} \ R}, \\
\label{S_d1}
\Sp & = & \SpExpr,
\end{eqnarray}

\noindent ignoring terms which do not depend on derivatives of the metric. In order for the variation principle to correctly yield Einstein's Equations \cite{York,GH}, we either need to fix $N_{ab} = K_{ab} - K \gal{ab}$ while varying $S_H$ or to fix $\gal{ab}$ while varying $S_H + \Sp$. We also can vary the difference $\Sp$ separately, with respect to both $\gal{ab}$ and $N_{ab}$ on $\partial V$ (but neither $g_{\alpha\beta}$ nor the bulk is involved),

\begin{equation}
\label{delta_S_d1}
\delta \Sp =  0, \qquad 
\end{equation}

\noindent i.e. the Gibbons-Hawking-York term is subject to the variational principle.

\section{Multiple horizon thermodynamics}
\label{thermodynamics}

Following \cite{Dim_Red}, we consider a 2+1-dimensional microscopic concept, namely on the non-null boundary $\partial V$ of a compact space-time region $V$. We restrict ourselves to the gauge $g_{\perp k\ne \perp}$ ($\perp$ is the component normal to $\partial V$).  
The boundary term $\Sp$ (\ref{S_d1}) can also be written as \cite{Padmanabhan_2014}

\begin{equation}
\label{S_d1_P}
\Sp = \cfac{16} \int_{\partial V} N^c_{ab} \ f^{ab} \ d^3\Sigma_c, \qquad f^{ab} = \sqrt{-g} \ g^{ab}, \qquad \NdSEq,
\end{equation}

\noindent where $N^c_{ab}$ is the momentum conjugate to $f^{ab}$ and $d^3\Sigma_c$ is the 3-surface element covector. We divide $\partial V$ into rectangular cuboids $\DPV \subset B_A \subset \partial V$ with edge length $L_e$ for each edge $e$ so that $L_p \ll L_e$ and $L_e$ is smaller than the typical variation scale of $\gamma_{ab}$ and $K_{ab}$. $\Sp$ then reads:

\begin{equation}
\label{S_d1_Delta_dV}
\Sp = \sum_{\DPV \subset \partial V} \SpD  = \cfac{32} \ \sum_{\DPV \subset \partial V} \inte{\DPV}{\sqn{\gamma} \ \gah{ab} \ \Lm_\perp \gal{ab}}.
\end{equation}

\noindent We diagonalise $\gal{ab}$ via an orthogonal transformation which is constant across $\DPV$, and we rewrite $\gal{ab}$ with induced triads $e_{(a)}^I$. We then replace $(x^a)=(x^\perp,\xpara, x^3,x^4)$, where $\xpara$ is tangent to $\partial V$, by null coordinates $(\tilde{x}^a)=(\tilde{x}^1=u,\tilde{x}^2=v, x^3,x^4)$. We split the integration range in (\ref{S_d1_Delta_dV}) into an interval $\mathcal{I}=[\xpara_-,\xpara_+]$ and a piece of 2-surface $A(\xpara)$. The integral over $\mathcal{I}$ can be written as a sum over $N = (\xpara_+-\xpara_-) / \Delta \xpara$ strips, with strip width $\Delta \xpara$ and locations $\xpara_k=\xpara_-+k\Delta \xpara$, where $\Delta \xpara$ spatially resolves $\Lm_\perp \gal{ab}$.
It can be shown \cite{Mandrin_2017} that (\ref{S_d1_Delta_dV}) is equal to

\begin{eqnarray}
\label{S_d1DdV_foliated}
S_{\partial 1 \ \DPV} & = & \cface{16} \sumk \intc{A(\xpara_k)}{2}{\sqn{\gamma} \ \Delta \xpara \ \sigma}, \\
\label{eq:sigma_dxpara}
\Delta \xpara \ \sigma & = & \Delta u \ e^{(a)_u}_I \ \Lm_v e^I_{(a)_u} + \Delta v \ e^{(a)_v}_I \ \Lm_u e^I_{(a)_v},
\end{eqnarray}

\noindent where $A(\xpara_k)$ is the intersection of $\DPV$ with the ($x^3,x^4$)-surface at $\xpara=\xpara_k$, $e^I_{(a)_w}$ are the triads induced on the hypersurface $w =$ constant with $w=u,v$. Diagonalising in turn $e^I_{(a)_w}$ and changing back to metric notation yields

\begin{eqnarray}
& & \SpD = \cface{32} \sumk \intct{A(\xpara_k)}{2}{\bigg[ \sqn{\gal{(u)}}\Jac{||}{u} \ \Delta u \ \gah{ab}_{(u)} \ \Lm_v \gal{ab}^{(u)} + \sqn{\gal{(v)}}\Jac{||}{v} \ \Delta v \ \gah{ab}_{(v)} \ \Lm_u \gal{ab}^{(v)} \bigg]} \nline
& & = \cface{32} \sumk \bigg[ \int_{\Delta u} \intct{A(\xpara_k)}{3}{\sqn{\gal{(u)}}\Jac{||}{u} \ \gah{ab}_{(u)} \ \Lm_v \gal{ab}^{(u)}} 
\nline
\label{general_null}
& & + \int_{\Delta v} \intct{A(\xpara_k)}{3}{\sqn{\gal{(v)}}\Jac{||}{v} \ \gah{ab}_{(v)} \ \Lm_u \gal{ab}^{(v)}} \bigg],
\end{eqnarray}

\noindent where $\gal{ab}^{(w)}$ and $\gah{ab}_{(w)}$ are the metric and inverse metric induced on the hypersurface $w =$ constant, respectively. Equation (\ref{general_null}) expresses $\SpD$ with respect to multiple null surfaces (layers) which can be interpreted using 
horizon thermodynamics \cite{Jacobson, Padmanabhan_2014} on null hypersurfaces $\Sigma$, via \cite{Majhi_Padmanabhan}

\begin{equation}
\label{S_MP}
\cfac{16} \int_{\Sigma} N^c_{ab} \ f^{ab} \ d^3\Sigma_c = \int_{\Sigma} T \ s \ d\lambda \ d^2x,
\end{equation}

\noindent where $s$ is the surface density of the entropy $S_{BHJP}$ and $T$ the temperature perceived by the Rindler observer on the horizon with parameter $\lambda = \bu$ or $\bv$. Therefore, (\ref{general_null}) can be written as

\begin{eqnarray}
\SpD & = & \half \sumk \bigg[ \int_{\Sigma_{(u)}(\xpara_k)} \tT{u} \ s \ du \ d^2x + \int_{\Sigma_{(v)}(\xpara_k)} \tT{v} \ s \ dv \ d^2x \bigg] \nline
\label{multi_observers}
& = & \half \sumk \int_{A(\xpara_k)} \big[ T_u \ s \ \Delta u  + T_v \ s \ \Delta v \big] \ d^2x,
\end{eqnarray}

\noindent where $\tT{w} = \Jac{||}{w} \ T_w$ and $w = u,v$. The multiple observer interpretation of (\ref{multi_observers}) in terms of hidden information has been discussed in \cite{Mandrin_2017}, and this justifies that $\SpD$ be denoted as multiple observer entropy or m-entropy, in contrast to the single observer entropy $S_{BHJP}$.


\section{Statistical gravity}
\label{sec:statistics}

Because of the thermodynamic property of the boundary space, the quantum bits of the multiple layers can be interpreted as the building blocks of a statistical theory as follows. The cuboid $\DPV$ is foliated across the direction defined by an arbitrary Minkowski index vector $x^K$ into $N_K$ pieces of 2d-surfaces of aria $A_K$ restricted on $\DPV$. The number $N_K$ is given by the best resolution achievable classically. $A_K$ represents $n_K$ quantum bits with $n_q \ge 2$ different possible states $|q\rangle$ each, so that $\DPV$ counts

\begin{equation}
\label{eq:Omega}
\Omega_K(\DPV) = ({n_q}^{n_K})^{N_K} = \exp{(s_K)}
\end{equation}

\noindent microscopic states and the m-entropy per volume $\DPV$ ($s_K$) and m-entropy ($S$) are

\begin{equation}
\label{eq:NK_nK}
s_K = N_K n_K \ln{n_q}, \qquad S = \sum_{\DPV_m \subset \partial V} s_{Km} \DPV_m,
\end{equation}

\noindent while $\gamma_{ab}$ is obtained via $A_K \approx 4L_p^2n_K$ and $\gah{ab}\Lm_\perp \gal{ab} \approx s_K$. Further relations and a more detailed description can be found in \cite{Mandrin_2017}. As a generalisation, the microstates are described by states labeled $s^{(i)}_{(j)}$ for every $i$th out of $n_{Kj}$ quantum bits on every $jth$ out of $N_K$ layers: 

\begin{equation}
\label{eq:state_gen}
|(s^{(1)}_{(1)},s^{(2)}_{(1)},\ldots,s^{(n_{K1})}_{(1)}); (s^{(1)}_{(2)},\ldots,s^{(n_{K2})}_{(2)}); \ldots ; (s^{(1)}_{(N_K)},\ldots,s^{(n_{KN_K})}_{(N_K)}) \rangle.
\end{equation}

\noindent A classical geometry corresponds to the macroscopic state with the highest probability. More precisely, if we divide the space of $\gamma_{ab}$ into small cells of equal size defined by a range $\Delta\gamma_{ab}$, the classical solution of GR lies in the cell which contains the largest number of microstates, or equivalently, at the stationary point of the boundary term.



\end{document}